\documentstyle [aps,prd]{revtex}
\input epsf
\begin{document}

\newcommand{\alp}{$\alpha\,\,$}
\newcommand{\xe}{$x_e\;$}
\newcommand{\tdot}{$\dot{\tau}\;\,$}
\newcommand{\be}{\begin{equation}}
\newcommand{\ee}{\end{equation}}
\newcommand{\obh}{$\Omega_b h^2\;$}
\newcommand{\omh}{$\Omega_m h^2\;$}
\newcommand{\och}{$\Omega_c h^2\;$}		
\newcommand{\okh}{$\Omega_K h^2\;$}	
\newcommand{\olh}{$\Omega_\Lambda h^2\;$}	
\title{The Effect of Time Variation in the Higgs Vacuum Expectation Value on the
Cosmic Microwave Background}
\author{Jens Kujat$^{1}$ and Robert J. Scherrer,$^{1,2}$}
\address{$^1$Department of Physics, The Ohio State University,
Columbus, OH~~43210}
\address{$^2$Department of Astronomy, The Ohio State University,
Columbus, OH~~43210}
\date{\today}
\maketitle

\begin{abstract}
A time variation in the Higgs vacuum expectation value alters
the electron mass and thereby changes the
ionization history of the universe.
This change produces a measurable imprint on the
pattern of cosmic microwave background (CMB) fluctuations.
The nuclear masses and nuclear binding energies, as well as the
Fermi coupling constant, are also altered, with negligible
impact on the CMB.
We calculate the changes in the spectrum of the CMB fluctuations as a function
of the change in the electron mass
$m_e$.  We find that future CMB experiments could be sensitive
to $|\Delta m_e/m_e| \sim |\Delta G_F/G_F| \sim 10^{-2} - 10^{-3}$.
However, we also show that a change in $m_e$ is nearly, but
not exactly, degenerate with a change in the fine-structure constant
$\alpha$.  If both $m_e$
and $\alpha$ are time-varying, the corresponding CMB limits
are much weaker, particularly for $l < 1000$.

\end{abstract}

\section{INTRODUCTION}

The possibility that the fundamental constants of nature
are not, in fact, constant, but might vary with time has
long been an object of speculation by physicists \cite{Dirac}.
The fundamental constants which have received the greatest
attention in this regard are the coupling constants
which determine the interaction strengths of the fundamental
forces:
the gravitational constant $G$, the fine-structure constant
$\alpha$, and the coupling constants for the weak
and strong interactions.  It has recently been noted that
measurements of the cosmic microwave background (CMB) fluctuations
in the near future will sharply constrain the variation
of $\alpha$ at redshifts $\sim 1000$ \cite{hannestad,kaplinghat};
here we extend this analysis to the Fermi coupling constant,
through its dependence on the Higgs vacuum expectation value.

As emphasized by Dixit and Sher \cite{dixit} (see also
reference \cite{spergel}) the Fermi constant
is not a fundamental coupling constant; it is actually
independent of the gauge coupling constant and depends
directly on the Higgs vacuum expectation value $\langle \phi \rangle$:
specifically, $G_F \propto \langle \phi \rangle^{-2}$.
Hence, it is most
meaningful to discuss constraints on the time
variation of $\langle \phi \rangle$,
rather than $G_F$.  Furthermore, the possibility of a time-variation
in the vacuum expectation value of a field seems more plausible
than the time variation of a fundamental coupling constant.
(For more detailed arguments in favor of considering (spatial) variations
in $\langle \phi \rangle$, see reference \cite{barr}).

Constraints on the time variation of $G_F$ or $\langle \phi \rangle$
have been considered previously in references \cite{dixit,spergel,weak}.
As noted in reference \cite{dixit},
changing $\langle \phi \rangle$ has four main physical effects with
astrophysical
consequences:
$G_F$ changes, the electron mass $m_e$ changes,
and the nuclear masses and binding energies change.
All four
of these alter Big Bang nucleosynthesis, and requiring consistency
with the observed element abundances gives limits
of \cite{spergel}
$\Delta G_F/G_F < 20\%$ at a redshift on the order of $10^{10}$.
In contrast, only one effect
is relevant for the CMB spectrum:  the change in $m_e$.
The weak interactions have no relevance at the epoch
of recombination, while the effect
of changing the nuclear masses and binding
energies is negligible compared to the effect of altering $m_e$.
Hence, for the purposes of the CMB, we can treat a change
in the Higgs vacuum expectation value as equivalent to a change in
$m_e$ alone, where $m_e \propto \langle \phi \rangle$.

In the next section, we describe the changes in recombination produced
by a change in $m_e$ and show how the CMB fluctuation
spectrum is altered.  We also examine the degeneracy between
altering $m_e$ and changing the fine structure constant $\alpha$.
In Sec. III, we translate our results into limits on a
time-variation in $m_e$
and, therefore, on the variation of $\langle \phi \rangle$ and $G_F$.
We find that
the MAP and PLANCK experiments might be sensitive to variations
as small as $|\Delta m_e/m_e| \sim 10^{-2} - 10^{-3}$, although
the limits are much weaker if $\alpha$ is allowed to vary as well.

\section{Changes in the recombination scenario and the CMB}

As in references \cite{hannestad,kaplinghat}, we will assume
that the variation in $m_e$ is sufficiently small during
the process of
recombination that we need only consider the difference
between $m_e$ at recombination and $m_e$ today; i.e., we treat
$m_e$ as constant during recombination.
The electron mass $m_e$ changes the CMB fluctuations
because it enters into the expression
for the differential optical depth $\dot \tau$ of photons
due to Thomson scattering:
\be
\label{taudot}
\dot{\tau}=x_en_pc\sigma_T,
\ee
where $\sigma_T$ is the Thomson scattering cross-section,
$n_p$ is the number density of electrons (both free and bound)
and $x_e$ is the ionization fraction.
The Thomson cross section depends on $m_e$ through the relation
\be
\sigma_T=8\pi\alpha^2\hbar^2/3m_e^2c^2.\
\ee
The dependence of $x_e$ on $m_e$ is more complicated; it depends
on both the change in the binding energy of hydrogen:
\begin{equation}
B = \alpha^2 m_e c^2/2,
\label{bind}
\end{equation}
which is the dominant effect, and also on the change in the recombination
rates with $m_e$.  Note that $m_e$ and $\alpha$ enter into
the expressions for $B$ and $\sigma_T$ in different ways, so that
the effect of changing $m_e$ cannot be parametrized in a simple
way in terms of the effect of changing $\alpha$ (calculated
in references \cite{hannestad,kaplinghat}).  However, since
the change in $B$ dominates all other effects, we expect
significant degeneracy between the effect of changing $m_e$ and
the effect of changing $\alpha$.  Since a change in $m_e$
affects the same physical quantities as a change in $\alpha$,
our discussion will parallel that in reference \cite{kaplinghat}.

The ionization fraction $x_e$ is determined by the ionization equation for hydrogen~\cite{peebles,jones}:
\be 
-\frac{dx_e}{dt}={\cal C}\left[{\cal R}n_px_e^2-\beta(1-x_e)
\exp\left(-\frac{B_1-B_2}{kT}\right)\right], 
\label{ion}
\ee
where ${\cal R}$ is the recombination coefficient, $\beta$ is the ionization 
coefficient, $B_n$ is the binding energy of the $n^{th}$ hydrogen atomic level and 
$n_p$ is the sum of free protons and hydrogen atoms. The Peebles correction 
factor ${\cal C}$ accounts for the effect of
non-thermal Lyman-$\alpha$ resonance photons and is given by: 
\be 
{\cal C}=\frac{1+A}{1+A+C}=\frac{1+K\Lambda (1-x_e)}{1+K(\Lambda+\beta)(1-x_e)},
\label{peeb}
\ee
where
$K=H^{-1}n_pc^3/8\pi\nu_{12}^3\) ($\nu_{12}$ is the Lyman-\alp transition frequency), and
$\Lambda$ is the rate of decay of the 2s excited state to the ground state via 2 photons and scales as $m_e$~\cite{breit}.
Since $\nu_{12}$ scales as $m_e$, we have
$K \propto m_e^{-3}$.
The ionization and recombination coefficients are related by the principle of detailed balance:
\be
\beta={\cal R}\left(\frac{2\pi m_e kT}{h^2}\right)^{3/2}\exp\left(-\frac{B_2}{kT}\right),
\ee
and the recombination coefficient can be expressed as 
\be
{\cal R}={\sum_{n, \ell}}^\star
\frac{(2\ell+1)8\pi}{c^2}\left(\frac{kT}{2\pi m_e}\right)^{3/2} \exp\left(\frac{B_n}{kT}\right)\int_{B_n/kT}^{\infty}\frac{\sigma_{n\ell}\;y^2\,d\!y}{\exp(y)-1}, 
\ee
where $\sigma_{n\ell}$ is the ionization cross-section for the 
$(n, \ell)$ excited level of hydrogen \cite{boardman}. In the above, the asterisk on the summation indicates that the sum from $n=2$ to $\infty$ needs to be regulated.
The $m_e$ dependence of the ionization cross-section
is rather complicated, but can be written as \(\sigma_{n\ell}\sim m_e^{-2}f(h\nu/B_1)\), from which one can derive the following equation:
\be 
\frac{\partial {\cal R}(T)}{\partial m_e}=-\frac{1}{m_e}\left(2{\cal R}(T) + T\frac{\partial {\cal R}(T)}{\partial T}\right). 
\label{rt}
\ee
This equation allows us to relate the $m_e$ dependence of the recombination coefficient to
its temperature parametrization ${\cal R}(T)$, which can be approximated by a power law of the form ${\cal R}(T) \sim T^{-\xi}$.
Then a solution of equation (\ref{rt}) has the $m_e$ dependence ${\cal R} \propto m_e^{\xi-2}$.
As in reference \cite{kaplinghat} we will take $\xi = 0.7$, corresponding to power law ${\cal R}(T) \sim T^{-0.7}$. We are interested in small changes
in $m_e$, so that $m_e^\prime=m_e(1+\Delta_m)$ with $\Delta_m \ll 1$.
Now equation (\ref{ion}) including a change in $m_e$ can be written as:
\be 
-\frac{dx_e}{dt}={\cal C}^\prime \left[{\cal R}^\prime n_px_e^2-\beta^\prime(1-x_e)\exp\left(-\frac{B_1^\prime-B_2^\prime}{kT}\right)\right], 
\label{ion2}
\ee
with ${\cal R}^\prime={\cal R}(1+\Delta_m)^{\xi-2}$, the changed binding energies $B_n^\prime=B_n(1+\Delta_m)$,
\be
\beta^\prime=\beta(1+\Delta_m)^{\xi-1/2}\exp\left(-\frac{B_2\Delta_m}{kT}\right),
\ee
and the changes in the Peebles factor (equation \ref{peeb}).
We then integrated equations (\ref{ion2}) and (\ref{taudot}) using CMBFAST \cite{CMBFAST} to obtain the CMB fluctuation spectra for different values of $m_e$.

Fig. 1 shows the results for a change in $m_e$ of $\pm 5$\% for a standard cold dark matter model (SCDM) with $h=0.65$ and $\Omega_bh^2=0.02$.
There are two main effects, similar to what is seen for
a change in the fine-structure constant \cite{kaplinghat}.  First, an
increase in $m_e$ shifts the curves to the right (i.e., larger $l$ values) due to the
increase in the hydrogen binding energy, which results in earlier recombination,
corresponding to a smaller sound horizon at the surface of last
scattering.  Second, the amplitude of the curves increases with
increasing $m_e$.  This second
change is due to two different physical effects:  an increase in the early
ISW effect due to earlier recombination (which
dominates at small $l$) and a change in the diffusion damping (which dominates
at large $l$) \cite{kaplinghat}.

\begin{figure}[Fig1]
\centering
\leavevmode\epsfxsize=12cm \epsfbox{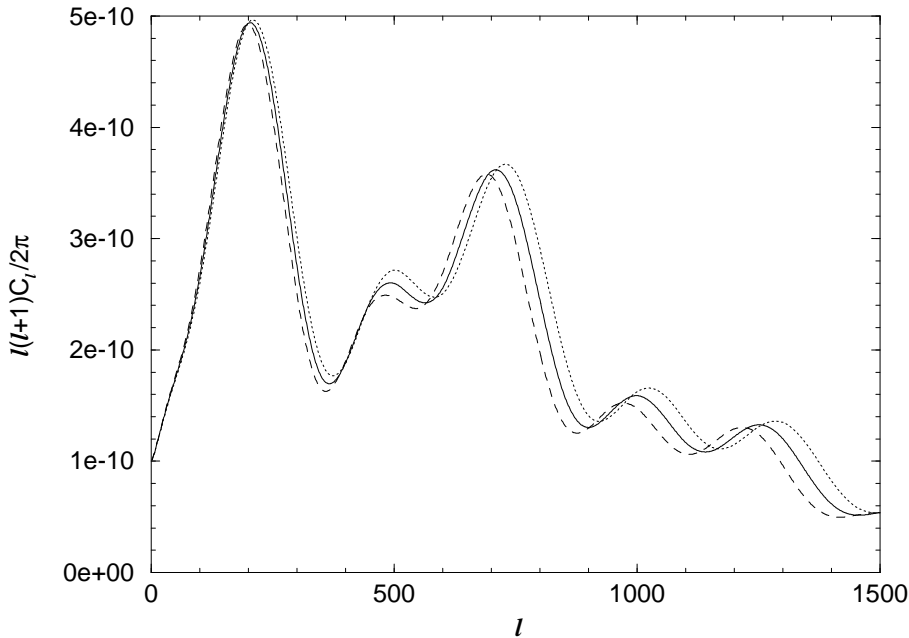}\\
\
\caption[Fig1]{\label{fig1}Spectrum of CMB fluctuations for a standard cold dark matter
scenario (SCDM, $\Omega_b h^2=0.02$, $h=0.65$) for no change in the electron
mass (solid curve) and for a
change in $m_e$ of $\pm 5\%$: $\Delta_m=+5\%$ (dotted curve) and
$\Delta_m=-5\%$ (dashed curve). }
\end{figure}

Since a change in $\alpha$ affects the same physical quantities as
a change in $m_e$, it is not surprising
that the effects on the CMB fluctuation spectrum are similar.  However,
they are not identical.  This can best be illustrated by
choosing changes in $m_e$ and $\alpha$ which leave the binding
energy $B$ unchanged, i.e., $(1 + \Delta_\alpha)^2 (1 + \Delta_m) = 1$,
since the change in $B$ dominates the changes in the fluctuation spectrum.
This is illustrated in Fig. 2, in which we have taken
a $3\%$ increase in $\alpha$ and a $5.74\%$ decrease in $m_e$.

\begin{figure}[Fig2]
\centering
\leavevmode\epsfxsize=12cm \epsfbox{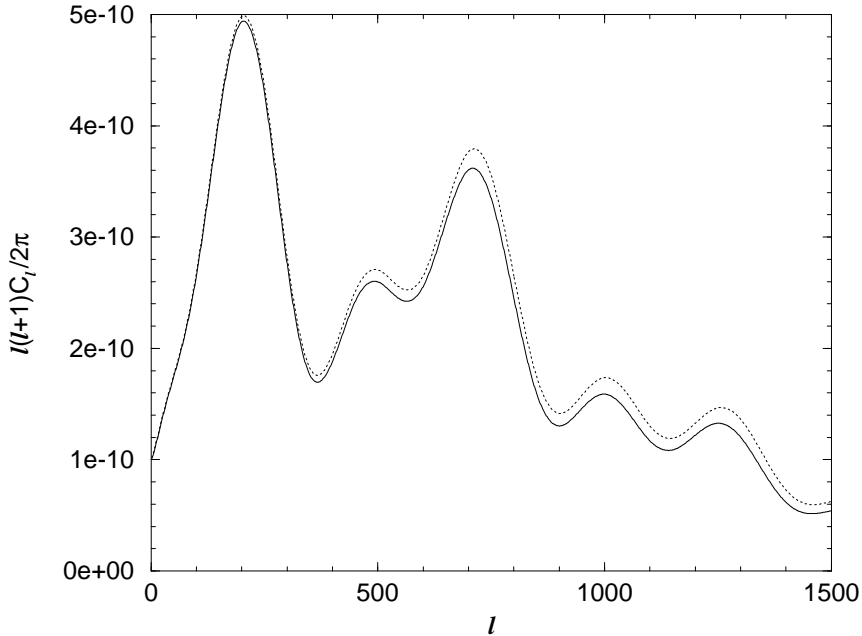}\\
\
\caption[Fig2]{\label{fig2}Spectrum of CMB fluctuations for a standard
CDM model
(SCDM,  $\Omega_b h^2=0.02$, $h=0.65$) (solid curve) and for
a change in both $\alpha$ and $m_e$, $\Delta_\alpha=+3\%$ and $\Delta_m=-5.74\%$,
which leaves the hydrogen binding energy unchanged (dotted 
curve).}
\end{figure}

As expected, the changes in $m_e$ and $\alpha$ nearly cancel in their
effect on the CMB, and there is no shift in the location of the peaks.
However, there is a residual increase in the amplitude
which is largest
at large $l$.  Recall that the shift in the position of the peaks
and the change in their amplitude at small $l$ are dominated by
the change in the binding energy, which is zero in this case.  However,
the change in the diffusion damping, which dominates the change in the
amplitude at large $l$, scales differently with $m_e$ and $\alpha$,
producing an increase in the peak amplitude at large $l$.
If both $\alpha$ and $m_e$ are assumed to be variable,
any CMB constraints on this variation will be considerably weaker.
There is some theoretical justification to consider such models
\cite{dixit,weak}.

\section{LIMITS ON VARIATIONS IN THE ELECTRON MASS}

We know from the analysis in references \cite{hannestad,kaplinghat} and in the previous
section that variations in $\alpha$ and/or $m_e$ will change the CMB spectrum significantly. In order
to impose limits on this variation from future CMB 
data, the Fisher information matrix is a very useful tool. 
For small variations in the parameters ($\theta_i$) of a cosmological model the likelihood function (${\cal L}$)
can be expanded about its maximum as
\be 
{\cal L}\simeq{\cal L}_m\exp(-F_{ij}\delta\theta_i\delta\theta_j), 
\ee  
where $F_{ij}$ is the Fisher information matrix, as defined in reference
\cite{tegmark}
\be 
F_{ij}=\sum_{\ell=2}^{\ell_{\mathrm max}}\frac{1}{\Delta{\cal C}_\ell^2}
\left(\frac{\partial{\cal C}_\ell}{\partial\theta_i}\right)
\left(\frac{\partial{\cal C}_\ell}{\partial\theta_j}\right),
\label{fij}
\ee
where
$\Delta{\cal C}_\ell$ is the error in the measurement of
${\cal C}_\ell$. In this approximation the inverse of the Fisher matrix $F^{-1}$ is the
covariance matrix, and in particular the variance of parameter $\theta_i$ is given by 
$\sigma_i^2=(F^{-1})_{ii}$. In the case of the CMB the cosmological parameters
($\theta_i$) that are taken to be determined from the measured fluctuation spectrum
are the Hubble parameter $h$,
the number density of baryons (parametrized as $\Omega_bh^2$), the cosmological
constant (parametrized as $\Omega_\Lambda h^2$), the effective number of relativistic
neutrino species $N_\nu$, and the primordial
helium abundance $Y_p$. Additionally, we allow the electron mass $m_e$
to serve as an undetermined parameter, and consider also the effect of adding
the fine-structure constant $\alpha$ to this set.
We make the assumption that the experiments are
limited only by the cosmic variance up to a maximum $\ell$, denoted by $\ell_{max}$.
Analysis of the Fisher information matrix will now enable us to
calculate a rough upper bound for the limits on $\Delta_m$ which
could be obtained from future CMB experiments.

We analyze two flat ($\Omega = 1$)
cold dark matter models,
a standard cold dark matter model
(SCDM) and one with $\Omega_\Lambda=0.7$ ($\Lambda$CDM).
Both models have $h=0.65$, $\Omega_bh^2=0.02$, $N_\nu=3.04$
and $Y_p=0.246$. For each model we calculate the variation in the electron
mass, $\sigma_m/m$, as a function of $\ell_{max}$ for
two different cases. In the first case we consider only $m_e$ to vary, taking
$\alpha$ as constant;
in the second case we take both $m_e$ and $\alpha$ to be variable.
The results are shown in Fig. 3. 
\begin{figure}[Fig3]
\begin{center}
\leavevmode
\vspace{-0.5cm}
\centerline{\epsfxsize=9cm \epsfbox{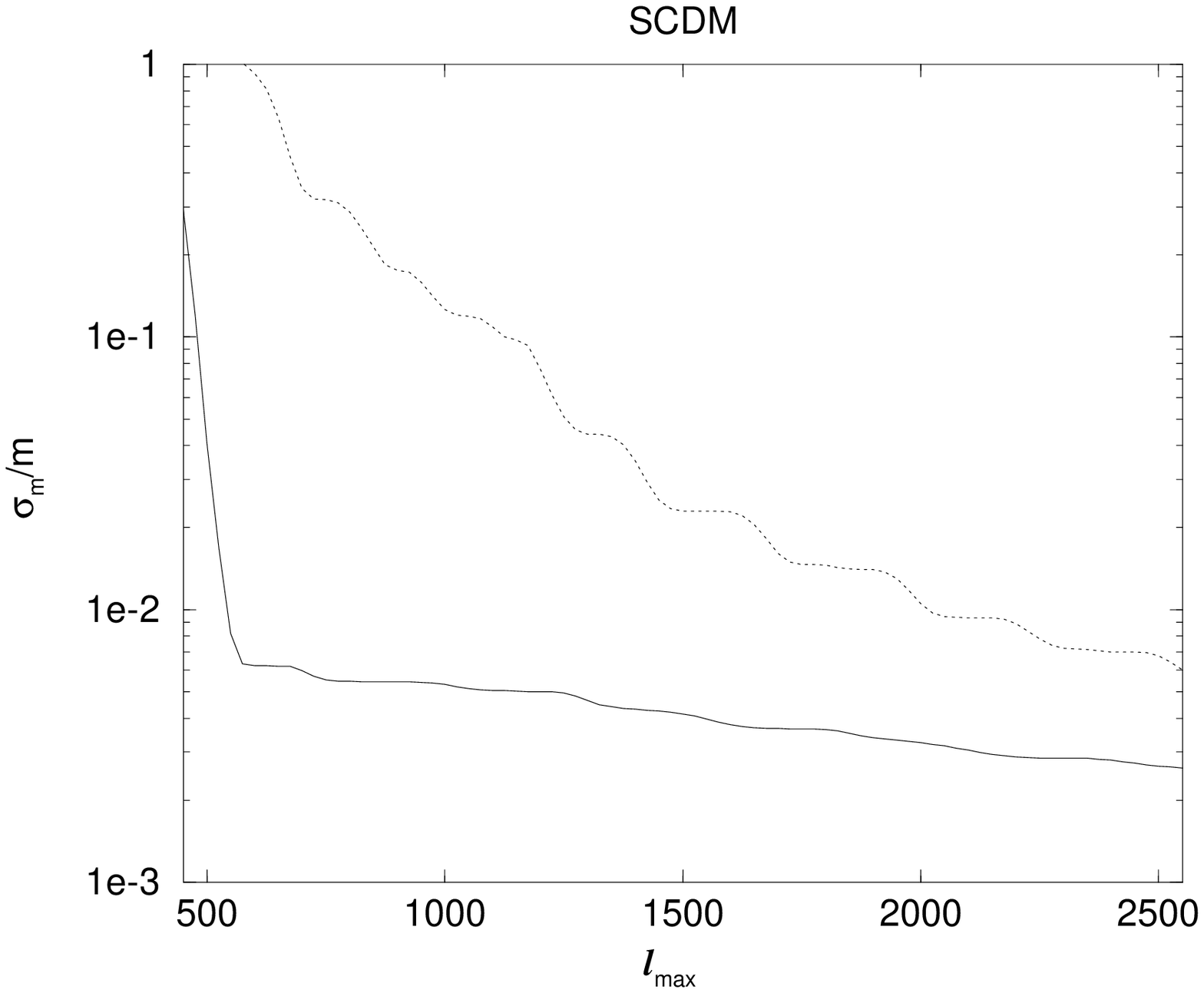} \epsfxsize=9cm \epsfbox{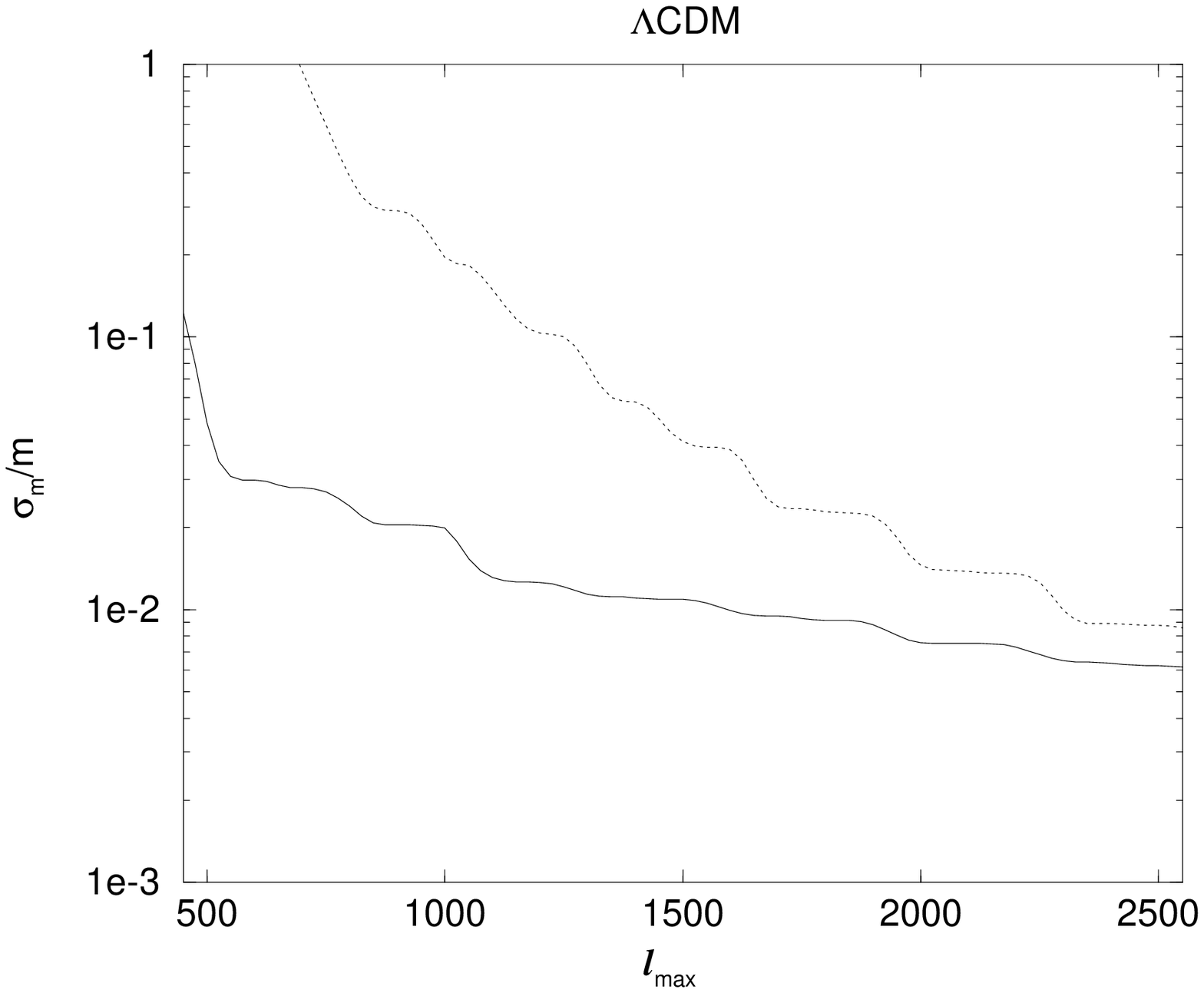}}
\end{center}
\
\caption[Fig3]{\label{fig3}The estimated accuracy with
which variations in $m_e$ could be constrained
by a future CMB experiment, as a function of the maximum angular resolution
given by $\ell_{max}$ for two cases: $m_e$ variable, but $\alpha$ constant
(solid curves), and both $m_e$ and $\alpha$ variable (dotted curves).}
\end{figure}

If $\alpha$ is taken to be constant, the upper limits on $|\Delta m_e/m_e|$ are
of order $10^{-2} - 10^{-3}$ for $\ell_{max} \sim 500 - 2500$ in both the
SCDM and $\Lambda$CDM models.
Since $m_e \propto \langle \phi \rangle$, and $G_F \propto \langle \phi \rangle^{-2}$,
similar limits apply to the variation in $\langle \phi \rangle$ and
$G_F$.  This represents potentially a much tighter limit on the time
variation in $G_F$ than can be obtained from Big Bang nucleosynthesis \cite{spergel}.
However, if we allow for an independent variation in both $m_e$ and
$\alpha$, then these limits become much less restrictive, since
these two effects are nearly degenerate.
For $\ell_{max} \sim 500 - 1000$ the limit on $|\Delta m_e/m_e|$ is
no better than 10\%,
while for $\ell_{max} > 1500$ it can be as small as $10^{-2}$.
This is consistent with the results shown in Fig. 2:  the degeneracy
between the effect of changing $m_e$ and the effect
of changing $\alpha$ is broken only
at the largest values of $l$.  As we have noted, there are models
in which simultaneous variation of $\langle \phi \rangle$ and
$\alpha$ occurs ``naturally" \cite{dixit,weak}.  Hence, our result
also supplies an important caveat to the limits on $|\Delta \alpha/\alpha|$
discussed in references \cite{hannestad,kaplinghat}:  these limits will apply
only if the Higgs vacuum expectation value is taken to be constant.

\vskip 0.2in
\noindent  We are grateful to M. Kaplinghat for helpful discussions,
and to S. Hannestad for useful comments on the manuscript.
We thank U. Seljak and M. Zaldariagga for the use of CMBFAST \cite{CMBFAST}.
This work was supported in part by the DOE (DE-FG02-91ER40690).

\end{document}